# Experimental evidence of giant pure optical activity in a metasurface based on a complementary twisted cross configuration

Ana Díaz-Rubio[1], Ben Tremain[2], Jorge Carbonell[1], José Sánchez-Dehesa[1] and Alastair Hibbins[2].

[1] *Wave Phenomena Group, Department of Electronics Engineering, Universitat Politècnica de València, Camino de Vera S/N. (Edificio 7F1), ES-46022 Valencia, Spain.*

[2] *Electromagnetic Materials Group, School of Physics, University of Exeter, Stocker Road, Exeter, EX4 4QL, United Kingdom*

**Abstract:** This work presents an experimental study of giant and pure optical activity in a periodic structure consisting of twisted crosses and complementary crosses patterned on the sides of a copper coated dielectric board. Additionally, a multilayer system is proposed and numerically studied to broaden the transmission bandwidth. Our results show that a dual band behavior can be obtained due to coupling effects between the layers whilst maintaining the dispersionless giant optical activity and negligible circular dichroism. We theoretically study the effect of the separation between layers and its influence on the transmission spectra.

Optical activity (the rotation of linear polarized radiation as it transmits through a material) is a phenomenon with important applications in analytical chemistry, biology, and crystallography [1]. In natural chiral materials, the optical activity is weak so, in order to obtain strong rotation angles, it is necessary to increase the physical dimensions of the chiral material. New possibilities were opened by using metamaterials; artificial structures made of subwavelength elements or meta-atoms. Their behavior under illumination can be described in terms of their constitutive parameters extracted using homogenization methods. These periodic structures allow one to tailor these parameters to obtain exotic properties.

Particularly, chiral metamaterials (CMs) have attracted a lot of attention as an alternative to obtain negative index media [2]. Also, it was demonstrated that CMs can be used for the design of perfect lenses [3]. Several double layer structures exhibiting optical activity have been proposed with rosettes, gammadions and U-shaped split resonators [1],[5],[6]. One of the simplest cases is a double layer system with twisted crosses which shows strong optical activity and negative refractive index [7]-10]. Using these CM layers, a bulk chiral metamaterial can be made simply by stacking them periodically. It has been demonstrated that the interaction between the CM layers modifies the properties of the resulting bulk material and one cannot predict the chiral behavior straightforwardly from the single layer behavior. However, these cases present a drawback, the resonant modes lead to a highly dispersive optical activity behavior.

A recent study of twisted crosses coupled to their complementary crosses shows strong dispersionless optical activity. Hannam *et. al.* demonstrated numerically and experimentally the optical activity (ϕ = 22°) of a cross and its complementary structure embedded in a circular waveguide [13][11]. However, Zhu *et.al.* theoretically analyzed an array of twisted crosses and complementary crosses obtaining enhanced optical activity of $\phi = 164°$ due to the strong interaction between nearest-neighboring meta-atoms.

In this letter, we report an experimental demonstration of giant optical activity in this array system. We also describe a proposal for improving the transmission bandwidth of the system which is obtained by means of a bulk chiral material made with two stacked layers of the cross - complementary cross metasurface. A comprehensive study of the coupling between layers is done through an eigenmode analysis.

1. **EXPERIMENTAL RESULTS**

The schematic view of a meta-atom of the chiral structure under study is shown in Fig.1(a) which is similar to that reported in [12]. The metasurface consists of a copper cross and a rotated complementary cross etched each on one side of a lossy dielectric sheet and placed in a square lattice. The thickness of the dielectric is $100\mu m$ with $\varepsilon_d = 2.7(1 + 0.02i)$. The lattice spacing is $d = 7.5mm$, the length and the width of the crosses are $l_{cross} = 6.25$mm and $w_{cross} = 375mm$; and the rotation angle between the cross and the complementary cross is $\theta = 22.5°$. The sample is formed by 50 unit cells in the x-direction and 37 in the y-direction.

We obtain the circularly polarized transmission coefficients by measuring the four linear transmission coefficients ($S_{xx}$, $S_{yx}$, $S_{yy}$ and $S_{xy}$). In the experimental setup the linearly polarized microwave radiation impinges perpendicularly on the sample from a rectangular waveguide horn antenna which is in the focus of a collimating parabolic mirror to generate planar waves. The transmitted beam is collected by a second rectangular horn placed at the focus of a second mirror. Both antennas are connected to a Vector Network Analyzer. The antennas can be azimuthally rotated 90° to allow the orthogonal polarizations needed in the experiment. The circular transmission coefficients for each polarization can be calculated from the linear transmission coefficients [11] as:

$$T_p = \frac{S_{xx} + S_{yy} \pm i(S_{xy} - S_{yx})}{2}, \qquad (1)$$

where p represents the left (LCP) or right (RCP) circular polarizations, which correspond to the + or − signs in the expression respectively. The optical activity, $\phi$, and the ellipticity, $\eta$, are defined as:

$$\phi = \frac{\arg(T_{RCP}) - \arg(T_{LCP})}{2}, \qquad (2)$$

$$\eta = \frac{|T_{RCP}|^2 - |T_{LCP}|^2}{|T_{RCP}|^2 + |T_{LCP}|^2}. \tag{3}$$

The experimental results are summarized in Fig.2. In Figure 2(a) the experimental amplitudes of the left and right circular polarization transmission coefficients (symbols) are compared with a numerical simulation which is represented with the red continuous line. Note that there is only one simulated result because left and right circular polarizations are equal. The simulations are performed using HFSS with one unit cell with periodic conditions in the *x* and *y* directions. In the *z* direction Floquet ports are located and the transmission coefficients are obtained from the S-parameters of these ports. A good agreement between theory and experiment is clearly observed, the maximum transmission amplitude is 0.61 at $30 GHz$. The experimental data also shows a second maximum at $34 GHz$ which is an effect of misalignment between the crosses and the complementary crosses. Figure 2(b) and 2(c) show the experimental and theoretical results for the optical activity and the ellipticity calculated with expressions (2) and (3) respectively. The behavior at frequencies between $18 GHz$ and $35 GHz$ is similar for the experimental and numerical results except for the frequencies surrounding $34 GHz$ due to the misalignment. The optical activity in the flat band is $\phi = 171°$ and the ellipticity at these frequencies is negligible ($\eta < 0.08$). Although the optical activity presents a broadband behavior, the useful bandwidth of the system is limited to the transmission bandwidth. A full width at half maximum calculation yields a fractional bandwidth of 7.4% for our structure

## 2. ENHANCEMENT OF THE TRANSMISSION BAND WIDTH

As we mentioned in the previous section the usable bandwidth of the system is restricted to the bandwidth of transmission. To solve this drawback and in order to increase the frequency range in which the system presents high transmission, we propose a system consisting of two stacked layers of these metasurfaces. Figure 1(b) shows the schematic for a stack of two layers of the cross-complementary cross metasurfaces. The air gap between both layers along $\hat{z}$ is denoted by the parameter $g$. The value of the parameter $g$ has to be small enough to ensure that the two CM are under the strong coupling regime.

In order to study the coupling of two of these chiral structures and the effects on the optical activity, we have numerically studied different configurations of the structure by changing the $g$ parameter. Figure 3 summarizes the results extracted from these simulations. Figure 3(a) shows the transmission coefficients for three different air gaps. Each transmission coefficient represents the transmission for both left and right circular polarizations, which are equal. We can see how the two layer system has a dual band transmission behavior. Also, it shows how the frequency shift between peaks becomes smaller when the air gap increases, i.e., the coupling effect decreases. In Figure 3(b), the optical activities for the same values of $g$ are displayed. There is a small reduction of the total flat band in these cases in comparison with the single layer case (see Fig.2); but at the frequencies of interest (frequencies at which there is high transmission) the results show no reduction of the optical activity. These results imply that the two layer system increases the bandwidth and keeps constant the flat band in optical activity. We define the bandwidth as:

$$BW = \frac{f_{max} - f_{min}}{f_0}, \qquad (4)$$

where $f_0$ is the central frequency between the transmission peaks, $f_{min}$ is the lower frequency when the first peak decreases 50% and $f_{max}$ is the higher frequency when the second peak is reduced by 50%. In the inset of Fig.3(b), the bandwidth variation with the air gap is represented. The result shows how the bandwidth decreases when $g$ increases because of the reduction of the coupling effect, therefore allowing control over the bandwidth of the system by changing the separation between layers. An air gap with $g = 400 \mu m$ has a transmission bandwidth $BW = 13.4\%$ which means an increase of 80% over the single layer case. In all the cases studied in this analysis, the minimum transmission in the usable bandwidth is higher than 0.25. The transmission in the system can be augmented by using another dielectric with less losses.

In figure 3.c, the ellipticity values for the same values of the air gap is shown. The ellipticity also remains small ($|\eta| < 0.008$), so we maintain the pure optical activity phenomenon in the frequencies of interest. These results show how it is possible to control the transmission spectra by changing the frequency split with the $g$ parameter, increasing the operational bandwidth and retaining the same optical activity properties of the one layer case.

To determine the nature of the dual band behavior in the stacked chiral material, we have performed an eigenvalue analysis of the structure. For this simulation, we work with a one cell system with periodic conditions in the x and y direction but, in this case, the Floquet ports are removed and substituted by PML conditions. The results of this analysis demonstrate that each peak is associated with one resonance of the system. In the inset of the Fig.3(a) the dependence of the frequency position of these modes with the air gap between layers is shown. The dashed red line marks the frequency of the mode for the single layer case. The high frequency mode, which is called the bonding mode, decreases in frequency when the air gap increases tending to the one single layer frequency. However, the low frequency mode or anti-bonding mode increases in frequency when the air gap increases and also tends to the one layer case. That means that, as expected, when the air gap increases the coupling between the layers vanishes and the modes tend to the one layer case.

To prove that the system with 2-layers has its own resonant modes which result from the original mode of the single layer case the surface currents of the eigenmodes are compared in Fig. 4. First, in panel (a), the surface currents for the single layer case are represented. The cut planes are xy-planes located in the z positions where the cross and complementary cross are placed (for convenience are denoted by Cross and CCross). We can see that, in the Cross plane, two dipoles are excited each one in one of the cross arms. Also in the CCross plane two dipoles are exited but with opposite current direction than in the Cross plane. Due to this fact, two current loops are generated, i.e the effective magnetic field causes the polarization rotation. A similar analysis has been done for the double layer system. Particularly, the numerical analysis is done with an air gap of $g = 500 \mu m$ whose transmission spectrum is reported in Fig.3(a). The eigenvalue analysis predicts two resonances of the system at $f_{ant-bonding} = 27.67 GHz$ and $f_{bonding} = 31.14 GHz$. A representation of the surface currents in these resonances is given in Fig.4(b) and Fig.4(c). In each panel, the surface current distribution for one mode in four different planes is represented. In Figure 4(a), the surface currents for the even mode are represented. The surface current distributions are similar but the current directions are opposite between the Cross1 and Cross2 planes and also between C.Cross1 and C.Cross2 planes. However in Figure 4(b), the bonding mode, the current distribution and the current directions are similar in the Cross planes and in the C.Cross planes.

## 3. CONCLUSIONS

In summary, this work has reported an experimental demonstration of pure giant optical activity in a metasurface consisting of a square array of twisted crosses and complementary crosses patterned on a dielectric board. These results confirm the theoretical prediction by Zhu *et. al*. In addition, these chiral metasurfaces can be used to form a bulk chiral material. A numerical study of a stacked structure formed by two CM layers has demonstrated that, with the cross-complementary cross system, it is possible to obtain strong coupling between layers producing wider transmission bandwidth and maintaining the optical activity properties. We have also shown how it is possible to tailor this dual band behavior by changing the separation between layers.

**ACKNOWLEDGMENTS**

This work was partially supported by the Spanish Ministerio de Economia y Competitividad (MINECO) under contract TEC2010-19751 and the grant EEBB-I-1408331.

**REFERENCES**

[1] Rogacheva, A. V., Fedotov, V. A., Schwanecke, A. S., & Zheludev, N. I. Giant gyrotropy due to electromagnetic-field coupling in a bilayered chiral structure. Physical Review Letters, 97(17), 177401. (2006)

[2] Monzon, C., & Forester, D. W. Negative refraction and focusing of circularly polarized waves in optically active media. Physical Review Letters,95(12), 123904. (2005).

[3] J. B. Pendry, A chiral route to negative refraction, Science 306(5700), 1353–1355 (2004).

[4] Monzon, C., & Forester, D. W. Negative refraction and focusing of circularly polarized waves in optically active media. Physical review letters,95(12), 123904. (2005).

[5] Gao, W., & Tam, W. Y. Optical activities in complementary double layers of six-armed metallic gammadion structures. Journal of Optics, 13(1), 015101. (2011)

[6] Li, Z., Zhao, R., Koschny, T., Kafesaki, M., Alici, K. B., Colak, E.& Soukoulis, C. M.. Chiral metamaterials with negative refractive index based on four "U" split ring resonators. Applied Physics Letters, 97(8), 081901. (2010)

[7] Dong, J., Zhou, J., Koschny, T., & Soukoulis, C. Bi-layer cross chiral structure with strong optical activity and negative refractive index. Optics Express, 17(16), 14172-14179. (2009).

[8] Zhou, J., Dong, J., Wang, B., Koschny, T., Kafesaki, M., & Soukoulis, C. M. Negative refractive index due to chirality. Physical Review B, 79(12), 121104. (2009).

[9] Decker, M., Ruther, M., Kriegler, C. E., Zhou, J., Soukoulis, C. M., Linden, S., & Wegener, M. (2009). Strong optical activity from twisted-cross photonic metamaterials. Optics letters, 34(16), 2501-2503.

[10] Li, Z., Alici, K. B., Colak, E., & Ozbay, E. Complementary chiral metamaterials with giant optical activity and negative refractive index. Applied Physics Letters, 98(16), 161907. (2011).

[11] Hannam, K., Powell, D. A., Shadrivov, I. V., & Kivshar, Y. S. Dispersionless optical activity in metamaterials. Applied Physics Letters,102(20), 201121. (2013)

[12] Zhu, W., Rukhlenko, I. D., Huang, Y., Wen, G., & Premaratne, M. Wideband giant optical activity and negligible circular dichroism of near-infrared chiral metamaterial based on a complementary twisted configuration. Journal of Optics, 15(12), 125101. (2013)

[13] Hannam, K., Powell, D. A., Shadrivov, I. V., & Kivshar, Y. S. Broadband chiral metamaterials with large optical activity. Physical Review B, 89(12), 125105. (2014)

[14] Li, Z., Caglayan, H., Colak, E., Zhou, J., Soukoulis, C. M., & Ozbay, E. Coupling effect between two adjacent chiral structure layers. Optics express,18(6), 5375-5383. (2010)

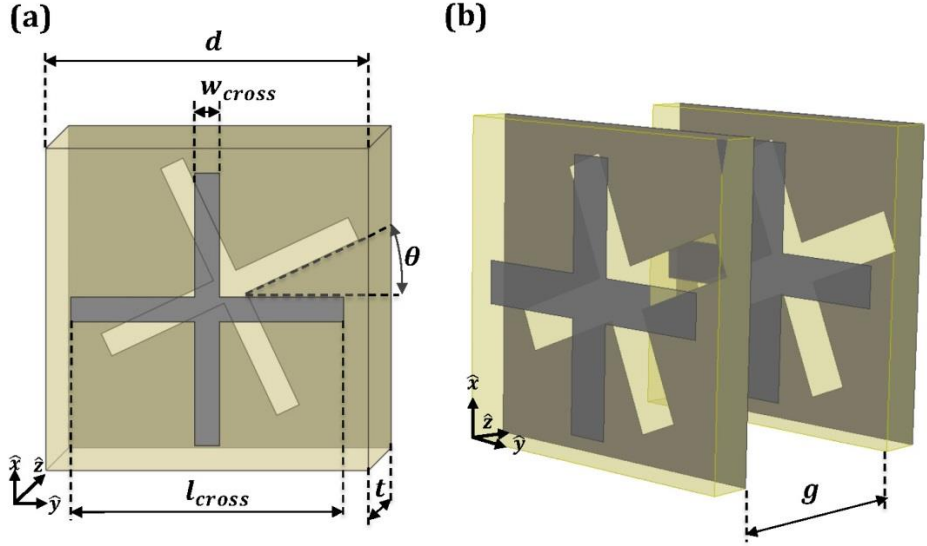

FIG. 1: Schematic representation of a meta-atom of the chiral surface. (a) Single layer and (b) double layer structure. The parameter $g$ describes the length of the air gap between layers in the z-direction.

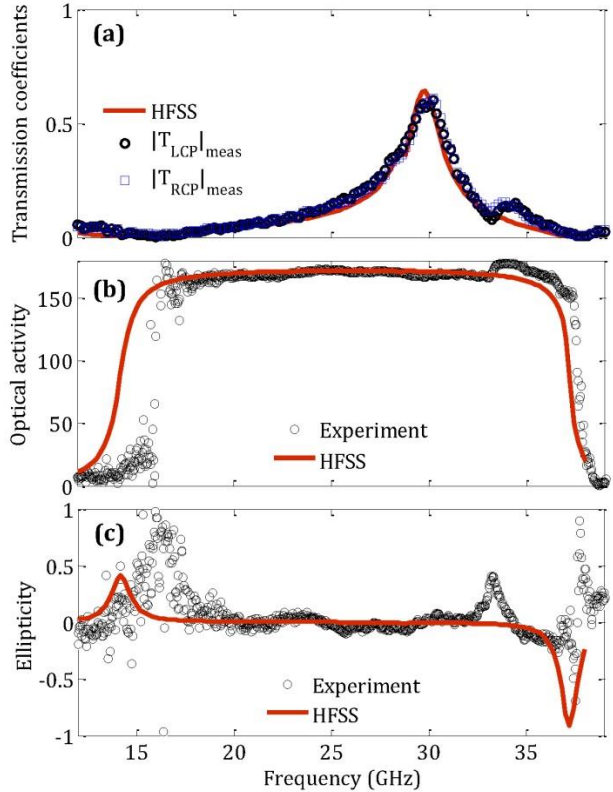

FIG. 2: Experimental demonstration of wideband giant optical activity on a cross-complementary cross system. (a) Circular transmission coefficients for the left and right circular polarization in comparison with the HFSS results. (b) Measured and simulated optical activity. (c) Measured and simulated ellipticity.

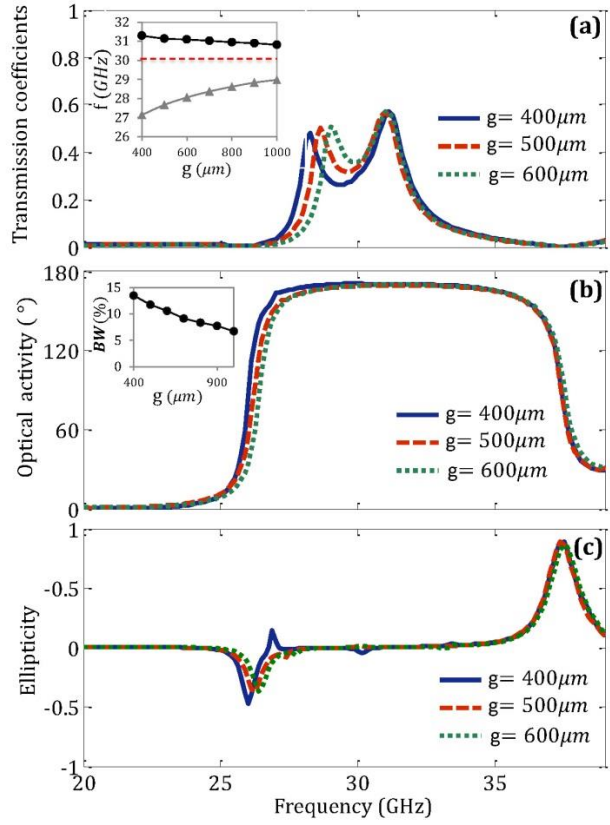

FIG. 3: Dual band giant optical activity on a multilayer cross- complementary cross system. (a) Circular polarization transmission coefficients and the inset represents the variation in frequency of both peaks as a function of the air gap. (b) Optical activity and the inset shows the dependence of the bandwidth with the air gap .(c) Ellipticity for different values of the air gap.

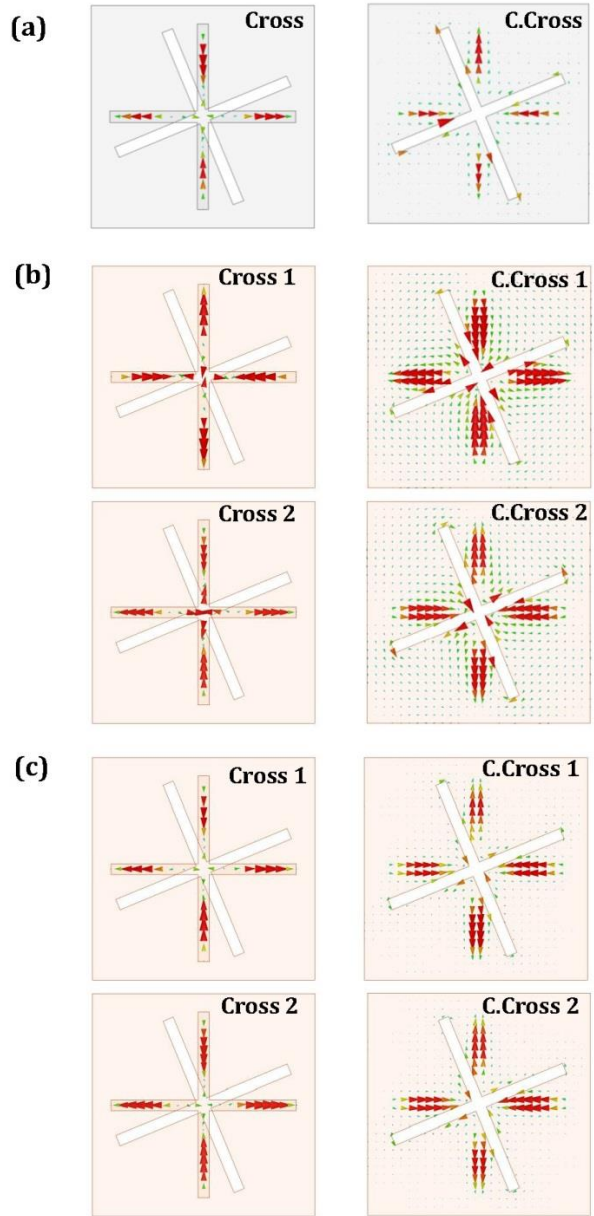

FIG. 4: (a) Surface currents in a single layer system. (b) and (c) surface currents in a two stacked metasurfaces with an air gap of $500\mu m$ between them:(b) the anti-bounding mode at $f = 27.67 GHz$ and (c) the bounding mode at $f = 31.14\ GHz$